\documentclass[%
 reprint,
 amsmath,amssymb,
 aps,
]{revtex4-1}

\usepackage{graphicx}
\usepackage{dcolumn}
\usepackage{bm}
\usepackage{siunitx}
\usepackage{float}
\usepackage{upgreek}
\usepackage{color}

\begin{document}

\preprint{APS/123-QED}

\title{Electromagnetically induced acoustic transparency with a superconducting circuit}

\author{Gustav Andersson}
 \email{gustav.andersson@chalmers.se}
\author{Maria K. Ekstr\"om}
\author{Per Delsing}
\affiliation{%
Department of Microtechnology and Nanoscience MC2, Chalmers University of Technology, Kemiv\"agen 9 SE-41296 G\"oteborg, Sweden
}%

\date{\today}

\begin{abstract}
We report the observation of Electromagnetically Induced Transparency (EIT) of a mechanical field, where a superconducting artificial atom is coupled to a 1D-transmission line for surface acoustic waves. An electromagnetic microwave drive is used as the control field, rendering the superconducting transmon qubit transparent to the acoustic probe beam. The strong frequency dependence of the acoustic coupling enables EIT in a ladder configuration due to the suppressed relaxation of the upper level. Our results show that superconducting circuits can be engineered to interact with acoustic fields in parameter regimes not readily accessible to purely electromagnetic systems. 

\end{abstract}

\maketitle

Electromagnetically induced transparency (EIT) is a quantum interference effect where an electromagnetic field controls the response of a three-level medium to a probe field \cite{Fleischhauer2005}. In contrast to the related phenomenon of Autler-Townes splitting, EIT arises due to the interference of excitation pathways in the coherent interaction of the atom(s) with the radiation field, and its signature is the appearance of a narrow transparent window inside the atomic absorption spectrum. EIT has been observed in atomic three-level systems with either $\Lambda$ or ladder-type configuration \cite{Gea1995,Gouraud2015}, and an analogue form of induced transparency has been demonstrated in optomechanical devices where light beams interact with a mechanical resonator through radiation pressure coupling \cite{Weis2010,Safavi-Naeini2011}. Due to the difficulty in engineering the requisite metastable states, EIT in a circuit quantum electrodynamics architecture \cite{Wallraff2004,Blais2004} was demonstrated only recently \cite{Novikov2015, Long2018}, using the combined states of an artificial atom in the form of a superconducting qubit and a three-dimensional microwave cavity. For superconducting circuits coupled to open transmission lines, what was thought to be observations of EIT \cite{Abdumalikov2010, Hoi2011} have been shown to in fact be the Autler-Townes splitting \cite{Anisimov2011}. 

Since its discovery, a range of potential applications for EIT in nonlinear optics and quantum information have been suggested, including quantum memories, routing and cross-phase modulation \cite{Ma2017,Xia2018,Schmidt96}. The modulation of absorption and emission cross-sections due to EIT could also have applications in the context of heat engines  \cite{Harris2016}. 

Strong coupling of superconducting qubits to surface acoustic waves (SAW) has been demonstrated in both waveguide and cavity settings \cite{Gustafsson2014, Moores2017}, and used to generate non-classical phonon states \cite{Satzinger2018} as well as for quantum state transfer \cite{Bienfait2019}. Here we exploit the strong frequency dependence of the piezoelectric coupling between a superconducting transmon circuit and SAW to control the relative decoherence rates of the transmon states, effectively enhancing the relative lifetime of the second excited state sufficiently to allow genuine EIT to occur. The delayline setup is commonly used to probe the properties of physical systems with SAW \cite{Wixforth1986,Weiler2012}  and provide an acoustic drive field, including demonstrations of coherent interference effects in optomechanics \cite{Balram2015}. Using a SAW delayline, we show electromagnetically induced acoustic transparency in both reflection and transmission of the probe. To our knowledge this constitutes the first observation of EIT for a propagating mechanical mode. For strong dressing fields the system crosses over to the Autler-Townes regime, where routing of SAW phonons has been shown by fast switching of the control tone \cite{ekstrm2019}.

\begin{figure*}
\begin{center}
\includegraphics[scale=0.9]{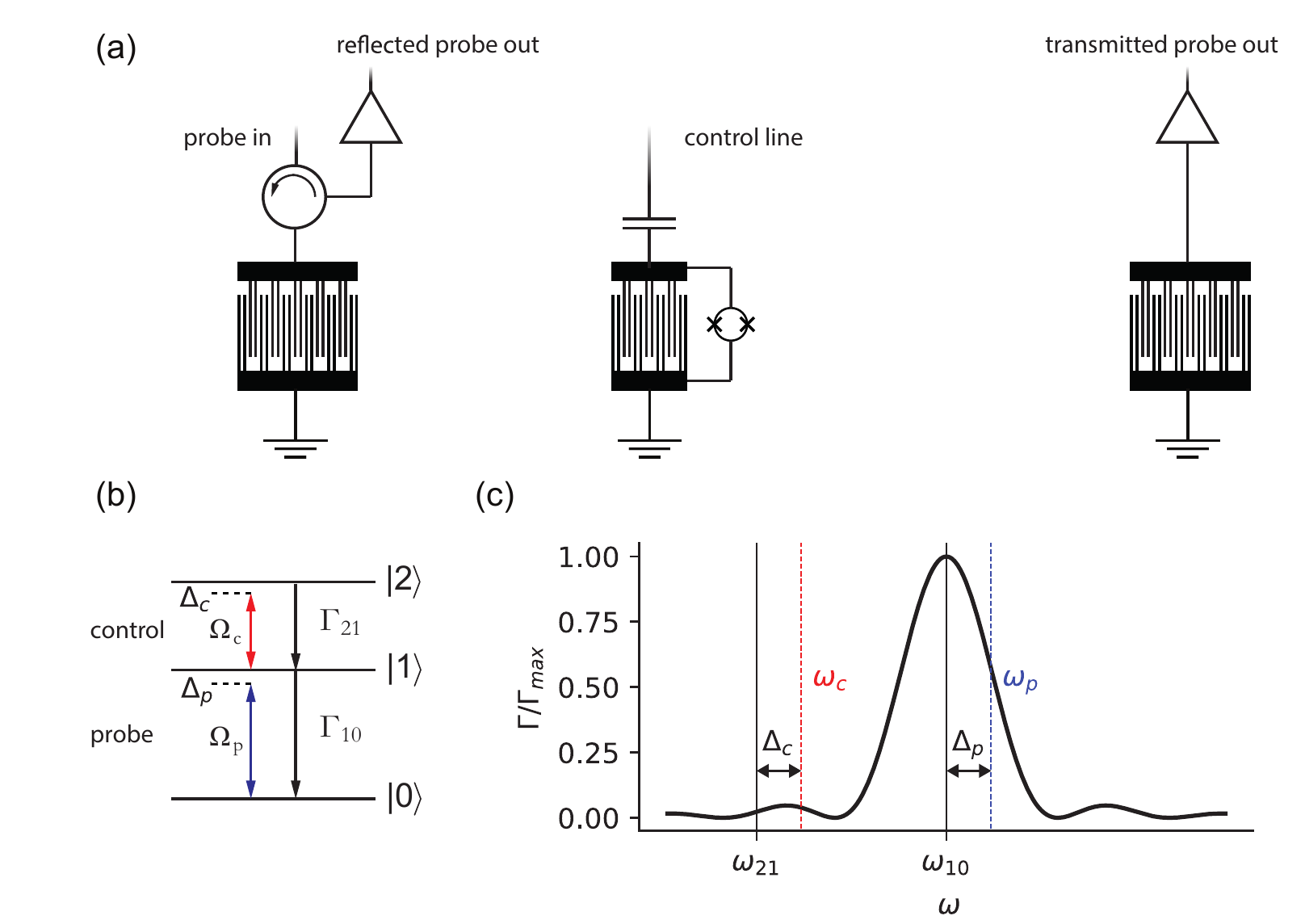}
\caption{\textbf{a}) Schematic of the device layout. An IDT with 150 finger pairs is used to launch the SAW probe beam towards the transmon and detect the reflected signal. The transmon cicruit interacts with the acoustic probe via its 25 IDT finger pairs, and the control field is applied to a gate electrode capacitively coupled to the transmon. A further IDT, also with 150 finger pairs, is used to receive and transduce the transmitted SAW signal. The structure is fabricated from aluminum on a piezoelectric GaAs substrate. \textbf{b}) Level scheme of the ladder system of the artificial atom. The acoustic probe field has an amplitude $\Omega_p$ and detuning $\Delta_p$ from the $\left| 0\right\rangle \leftrightarrow  \left| 1\right\rangle$ transition. A purely electromagnetic control field is applied close to the $\left| 1\right\rangle \leftrightarrow \left| 2\right\rangle$ transition with detuning $\Delta_c$. The direct $\left| 0\right\rangle \leftrightarrow \left| 2\right\rangle$ transition is suppressed by parity conservation and spontaneous emission occurs to the next-lower level. \textbf{c}) Schematic illustrating the spontaneous emission rate of the artificial atom into the SAW transmission line, which depends on frequency as the square of a sinc function [supplementary information]. We operate the device in a regime where the $\left| 0\right\rangle \leftrightarrow  \left| 1\right\rangle$ transition at $\omega_{01}$ is maximally coupled, and the $\left| 1\right\rangle \leftrightarrow \left| 2\right\rangle$ transition is completely outside the main lobe of the coupling function.}
\label{schematic}
\end{center}
\end{figure*}

The ladder-type three-level system is formed by the ground and two first excited states of a superconducting transmon circuit \cite{Koch2007}. The transmon, which is fabricated on a GaAs substrate, couples piezoelectrically to the propagating SAW field via an Interdigitated Transducer (IDT). The IDT spans $N_p=25$ periods and has a double-finger structure to suppress internal mechanical reflections \cite{Bristol1972}. A SQUID loop allows for tuning of the transition frequencies. Whereas a quantum emitter in an electromagnetic transmission line couples to all modes of the propagating field, the periodic structure of the IDT restricts the coupling of the SAW-qubit interaction to a bandwidth of approximately $\Delta_\textup{IDT}/2\pi = 0.9f_\textup{IDT}/N_p$ \cite{Datta1986, Aref2015}. With an IDT center frequency $f_\textup{IDT}$ of \SI{2.26}{GHz}, we obtain $\Delta_\textup{IDT}/2\pi \approx$ 81 MHz. The transition frequency $\omega_{10}$ is tuned into this band using an external magnetic flux. The IDT finger structure provides a capacitance to the transmon circuit giving rise to a charging energy of $E_C/h=\SI{129}{MHz}$. As the transmon anharmonicity, given by $E_C$, is sufficient to ensure $\Delta_\textup{IDT}<\omega_{10}-\omega_{12}$, maximizing the SAW coupling of the first transition by setting $\omega_{10}/2\pi=f_\textup{IDT}$ implies $\Gamma_{10} \gg \Gamma_{21}$ where $\Gamma_{ij}$ denotes the spontaneous emission rate from the state $i$ to state $j$. The filtering provided by the IDT thus strongly suppresses the coupling of higher transmon levels to SAW.

A weak SAW probe beam is launched towards the transmon using a double-finger IDT with 150 periods, located a distance 300 $\upmu$m away. The acoustic signal reflected back to the launcher IDT is measured using a microwave circulator while a control field is applied via a capacitively coupled electrical gate, as shown schematically in Fig.~\ref{schematic}. 
 
\begin{figure}[!htbp]
\begin{center}
\includegraphics[scale=0.85]{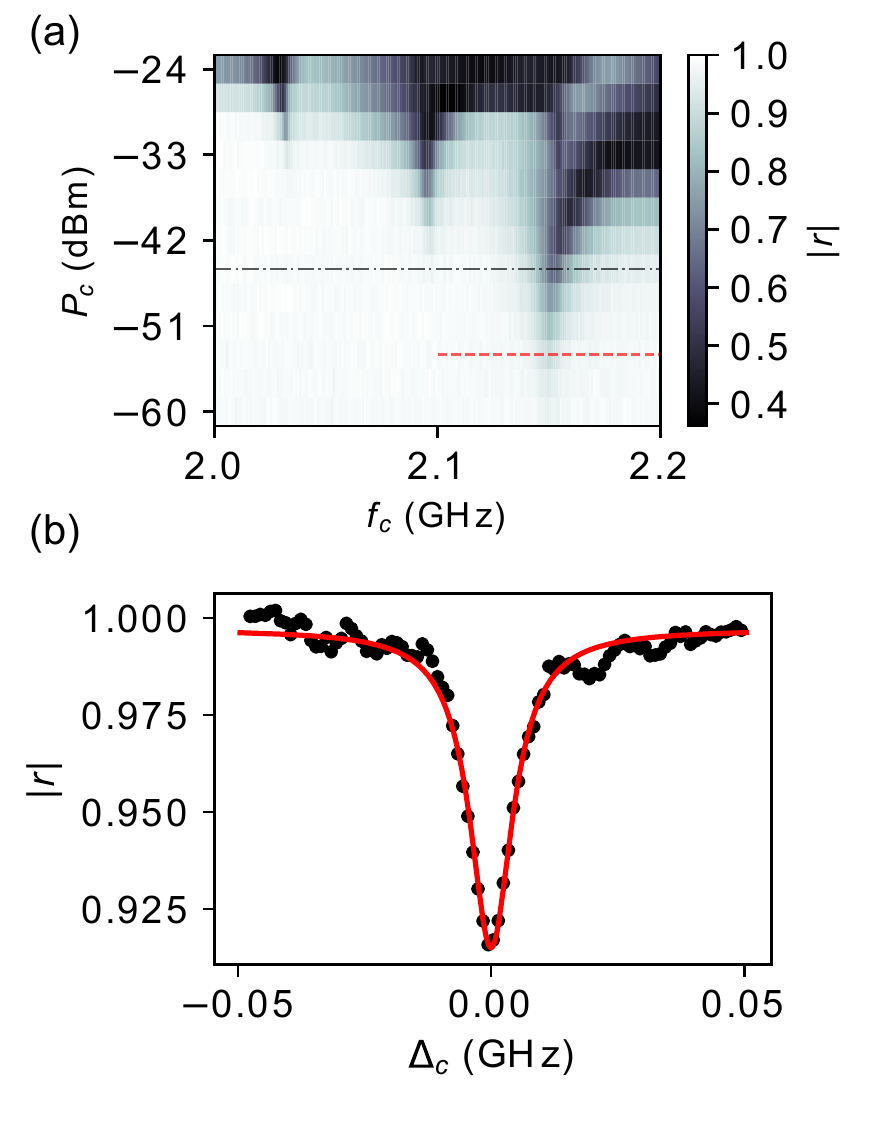}
\caption{\textbf{a} Normalized reflection coefficient measured as a function of (room temperature) control power $P_c$ and frequency of the control field. The probe frequency is \SI{2.2684}{GHz}. The black dotted line indicates the crossover from the low power EIT to the Autler-Townes effect. With increasing power the level scheme of Fig.~\ref{schematic} breaks down as multiphoton transitions are strongly driven. The red dashed line indicates the line cut shown with a lorentzian fit (red line) in \textbf{b}. The dip in reflection is due to the EIT.}
\label{reflection_pow}
\end{center}
\end{figure}

The reflection coefficient of the probe beam is given by \cite{Astafiev2010}
\begin{equation}
r=i\frac{\Gamma_{10}}{\Omega_p}\left\langle\sigma_-\right\rangle
\end{equation}
where $\Omega_p$ is the Rabi frequency given by the probe amplitude and $\sigma_-$ denotes the lowering operator between the $\left| 0\right\rangle$ and $\left| 1\right\rangle$ transmon states. For a weak probe under the application of a control field this yields in the steady state \cite{Abdumalikov2010}
\begin{equation}
r = -\frac{\Gamma_{10}}{2\left(\gamma_{10}-i \Delta_p\right)+\frac{\Omega_c^2}{2\left(\gamma_{20} - i \Delta_p - i \Delta_c\right)}}
\label{eq: reflection}
\end{equation}
where $\Delta_p=\omega_p-\omega_{10}$ is the probe detuning. The control field has an effective Rabi frequency $\Omega_c$ and a detuning $\Delta_c=\omega_c-\omega_{21}$. The decoherence rates $\gamma_{ij}$ correspond to decay rates of the off diagonal density matrix elements and determine the conditions for realizing EIT. A common procedure in optical EIT measurements is to sweep the probe detuning $\Delta_p$ while applying a resonant control field ($\Delta_c=0$) \cite{Fleischhauer2005}. The limited bandwidth of the probe IDT makes this approach impractical in our case. Instead, we adopt a reversed scheme where the frequency of the acoustic probe is fixed on resonance while sweeping the frequency and amplitude of the control field. As the control field interaction is purely electromagnetic it can be applied across a wide frequency range. Figure~\ref{reflection_pow} A shows the reflected probe amplitude as a function of applied control power and frequency for a probe beam at \SI{2.2684}{GHz}. As the control frequency is swept into resonance with the $\left| 1\right\rangle \leftrightarrow \left| 2\right\rangle$ transition at $f_{12}=\SI{2.15}{GHz}$, a dip appears in the reflected amplitude. Together with an independent estimate of $\gamma_{10}$, this measurement allows for extracting all parameters relevant to discerning EIT from the Autler-Townes splitting \cite{Anisimov2011}. The rate $\gamma_{10}$ is obtained from analyzing the lineshape obtained while sweeping the transmon frequency $\omega_{10}$ around the probe frequency in the absence of a control field ($\Omega_c=0$), and found to be $\gamma_{10}/2\pi=\SI{21}{MHz}$. 

\begin{figure}
\begin{center}
\includegraphics[scale=0.85]{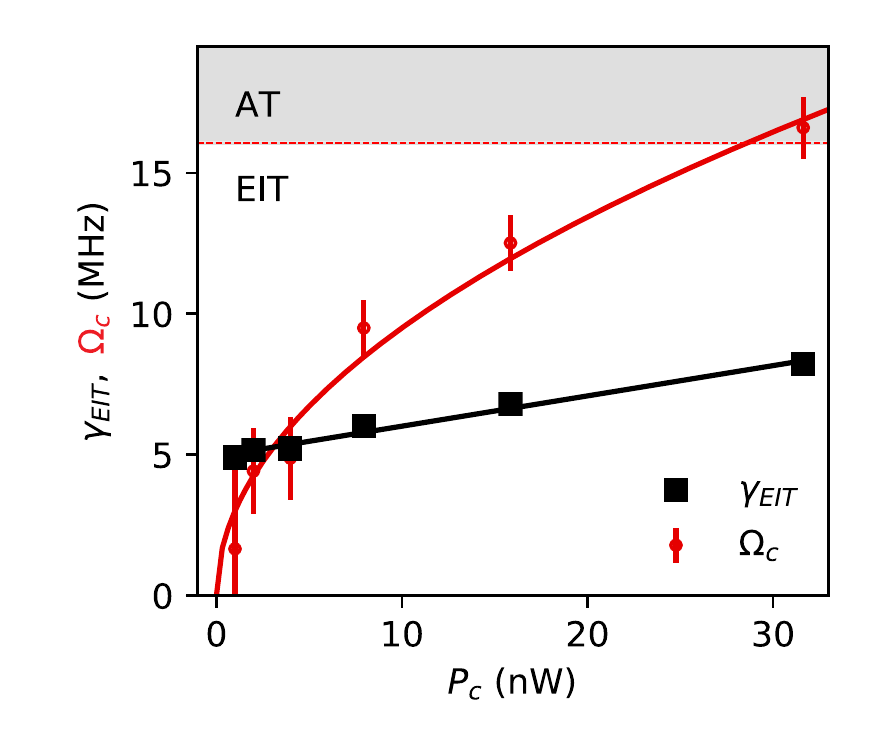}
\caption{EIT linewidth $\gamma_\textup{EIT}$ as a function of the power $P_c$ applied to the control gate at room temperature. The black line shows a linear fit. The extracted fit parameters are used to obtain $\gamma_{20}=\SI{4.94}{MHz}$ and the effective control strength $\Omega_c$ as a function of power (solid red line) from Eq.~\ref{eq: linewidth}. The red dots show the extracted $\Omega_c$ for each data point with error bars. The threshold $\Omega_c<\gamma_{10}-\gamma_{20}$ is indicated by the dashed red line.}
\label{linewidths}
\end{center}
\end{figure}

A resonant probe beam implies setting $\Delta_p=0$ in Eq.~\ref{eq: reflection}, which yields a negative Lorentzian with a half with at half maximum given by
\begin{equation}
\gamma_\mathrm{EIT} = \gamma_{20}+\frac{\Omega_c^2}{4\gamma_{10}}.
\label{eq: linewidth}
\end{equation}
This linewidth is linear in the control power and limited in sharpness by the decoherence $\gamma_{20}$ of the upper level in the ladder. In Fig.~\ref{linewidths} we plot the linewidth of this Lorentzian as a function of the control tone power applied at room temperature as well as linear fit. From the fit we extract a $\gamma_{20}/2\pi$ of \SI{4.94 \pm 0.14}{MHz}. As the coupling of the $\left| 1\right\rangle \leftrightarrow \left| 2\right\rangle$ transition to SAW is suppressed by more than one order of magnitude relative to $\left|1\right\rangle$, this decoherence rate is not dominated by SAW emission, but rather other sources of pure dephasing. We stress that the condition $\gamma_{10}>\gamma_{20}$ arises due to the frequency dependence of the acoustic coupling, and is necessary to enable EIT in ladder type systems. While the constant term of the linear fit yields $\gamma_{20}$, we use the known $\gamma_{10}$ and the slope to extract the control field drive strength $\Omega_c$ as a function of power applied at room temperature.

The presence of a transparent window in the scattering off a three-level system does not necessarily imply EIT and substantial theoretical analysis has been developed to determine whether EIT or Autler-Townes splitting occurs under given experimental conditions \cite{Anisimov2011}. EIT relies on the destructive interference of two excitation paths. In the ladder case the direct $\left| 0\right\rangle \rightarrow  \left| 1\right\rangle$ transition interferes with the path exciting to the upper level and back, $\left| 0\right\rangle \rightarrow  \left| 1\right\rangle \rightarrow  \left| 2\right\rangle \rightarrow  \left| 1\right\rangle$. Spontaneous emission or dephasing of the state $\left|2\right\rangle$ suppresses this interference. Using the criterion from \cite{Anisimov2011}, the quantitative distinction arises from analyzing the poles of Eq.~\ref{eq: reflection}. If under resonant control ($\Delta_c=0$), the poles of Eq.~\ref{eq: reflection} are purely imaginary, the reflection coefficient as a function of probe frequency can be expressed as the difference of two Lorentzians centered at the same frequency. This is the condition for EIT and in our system equivalent to $\Omega_c<\gamma_{10}-\gamma_{20}$, which implies EIT can only be observed if $\gamma_{10}>\gamma_{20}$. The threshold for the drive amplitude that separates the EIT and AT regimes is shown as the red dashed line in Fig.~\ref{linewidths}. If the drive strength is increased beyond this limit the reflection coefficient is given by the sum of two Lorentzians separated by the drive strength $\Omega_c$. This is the Autler-Townes splitting. As shown in Fig.~\ref{linewidths},  the lower control powers are insufficient for Autler-Townes splitting to appear and the transparency features observed are due to the EIT. The crossover to the Autler-Townes regime appears at $\Omega_{c,t}=\gamma_{10}-\gamma_{20} = 2\pi \cdot \SI{16.1}{MHz}$, corresponding to a control power of -45 dBm. Figure~\ref{reflection_pow} shows the dip in reflection arising from EIT at a drive amplitude of $\Omega_c/2\pi=\SI{6.1}{MHz}$, well below $\Omega_{c,t}$. 

\begin{figure}
\begin{center}
\includegraphics[scale=0.85]{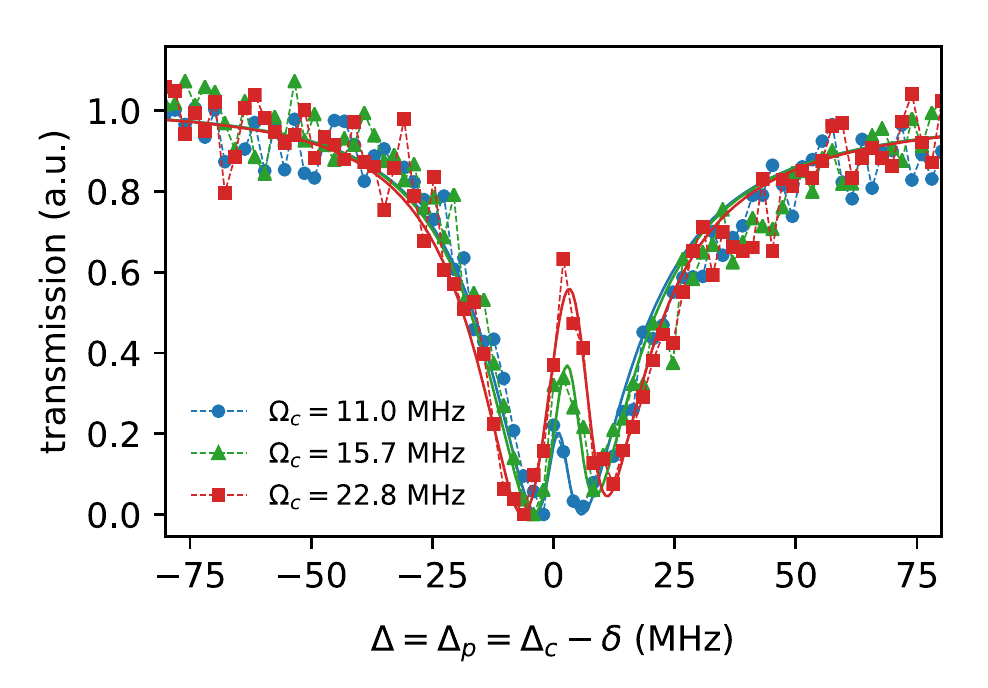}
\caption{Normalised transmission coefficient in the EIT (blue), crossover (green) and Autler-Townes (red) regimes. Solid lines are fits to data and yield $\gamma_{20}/2\pi=\SI{4.5}{MHz}$. The probe tone is fixed at 2.2644 GHz, and the probe and pump detunings are swept by tuning the transmon frequency via an external magnetic field. Due to the fixed anharmonicity of the transmon, the $\omega_{10}$ and $\omega_{21}$ transition frequencies are swept uniformly. The slight asymmetry in the lineshape is due to a residual control detuning $\delta\approx \SI{4}{MHz}$} at $\Delta_p=0$ as well as the interference of electromagnetic crosstalk between the launcher and receiver IDTs.
\label{transmissionfluxfreq}
\end{center}
\end{figure}

We further measure the acoustic EIT in transmission, where the transmitted SAW signal is transduced by an IDT at a distance 100 $\upmu$m from the transmon. In this measurement we use a different scheme where the probe frequency is fixed, and an external magnetic flux is used to tune the qubit frequency $\omega_{10}/2\pi$, thereby sweeping $\Delta_p$. As the control frequency also remains fixed, this scheme will simultaneously sweep both $\Delta_c$ and $\Delta_p$. To first order, the anharmonicity is not affected by the tuning, yielding $\Delta_c = \Delta_p + \delta$, where $\delta$ is the residual control detuning at probe resonance $\Delta_p=0$. The case $\delta=0$ corresponds to perfect alignment of control and probe frequencies with the anharmonicity. With the definition $t=1+r$ and $\Delta_p$ as the independent variable we get the transmission coefficient
\begin{equation}
t = 1-\frac{\Gamma_{10}}{2\left(\gamma_{10}-i \Delta_p\right)+\frac{\Omega_c^2}{2\left(\gamma_{20} - 2i \Delta_p - i \delta\right)}}.
\label{eq: transmission}
\end{equation}
Figure~\ref{transmissionfluxfreq} shows the normalised transmission amplitude for different drive strengths. We extract an upper level decoherence rate of $\gamma_{20}/2\pi=\SI{4.5 \pm 0.6}{MHz}$. While this estimate is less precise than the result obtained in the reflection measurement, they are consistent insofar as the error margin in reflection (\SI{4.94 \pm 0.14}{MHz}) falls completely within this range.

In conclusion, we have demonstrated that EIT can be generated in a mechanical mode of propagating surface acoustic waves by using an electromagnetic control signal. We show consistent data in reflection and transmission. By demonstrating quantum interference of acoustically and electromagnetically driven excitations, this experiment suggests applications in phononic quantum information architectures, where SAW phonons couple disparate quantum systems. We show that the  piezoelectric coupling of superconducting qubits to short-wavelength SAW has a frequency dependence offering the possibility to engineer relaxation rates, allowing EIT to be observed in waveguide QED with superconducting circuits. This principle could be further exploited in quantum acoustic experiments \cite{Kockum2014}, for example to generate population inversion and single-atom sound lasing. EIT is further associated with slow light \cite{Hau1999}, leading to a reduction in the group velocity as the field propagates through the EIT medium. The amount of slowdown is limited by the linewidth of the EIT window. While in our case this limit corresponds approximately to a factor of three ($v_g\approx \SI{1000}{m/s}$), improved coherence in the second excited state would enable larger reductions of the sound velocity. Larger group delays could also be achieved by using an array of artificial atoms rather than only a single transmon.

We acknowledge fruitful discussions with B. Suri and G. Johansson. This work was supported by the Knut and Alice Wallenberg foundation and by the Swedish Research Council, VR. This project has also received funding from the European Union's Horizon 2020 research and innovation programme under grant agreement No 642688 (SAWtrain).


\bibliography{EITbib}
\section*{Supplementary information}
\subsection*{Semiclassical model for the acoustic coupling}
The frequency dependence of the coupling strength of the qubit derives from the frequency response of the IDT.  Approximating the SAW-coupled qubit as a classical $RLC$ circuit dissipating energy stored in the $LC$ resonator by conversion to SAW, the conductance due to phonon emission is given by \cite{Datta1986, Aref2015}
\begin{equation}
    G_a = G_{a0}\left(\frac{\sin{X}}{X}\right)^2,
\end{equation}
where $X=N_p\pi\frac{\omega-\omega_\mathrm{IDT}}{\omega_\mathrm{IDT}}$ and $G_{a0} \approx K^2N_p\omega_\mathrm{IDT} C_t$. Here, $K^2$ is the electromechanical coupling coefficient and $C_t$ the total capacitance of the qubit. This gives a decay rate of the qubit into the SAW channel of
\begin{equation}
\Gamma_a = \frac{\omega_\mathrm{IDT}G_a}{2}\sqrt{\frac{L_J}{C_t}} = \frac{G_a}{2C_t}
\end{equation}
where $L_J$ is the SQUID inductance. With our expression for $G_a$ this yields
\begin{equation}
    \Gamma_a =\Gamma_{a0}\left(\frac{\sin{X}}{X}\right)^2,
    \label{eq. sinc}
\end{equation}
where $\Gamma_{a0}\approx 0.5K^2 N_p \omega_\mathrm{IDT}$. For our device we estimate a spontaneous emission rate on resonance ($\omega_\mathrm{IDT}=\omega_{10}$) of $\Gamma_{10}/2\pi=\Gamma_{a0}/2\pi=\SI{20.1}{MHz}$. The frequency dependence given by Eq.~\ref{eq. sinc} is shown in Fig.~\ref{schematic} (c) of the main text.

\subsection*{EIT linewidth estimation}
\begin{figure}
\begin{center}
\includegraphics[scale=0.85]{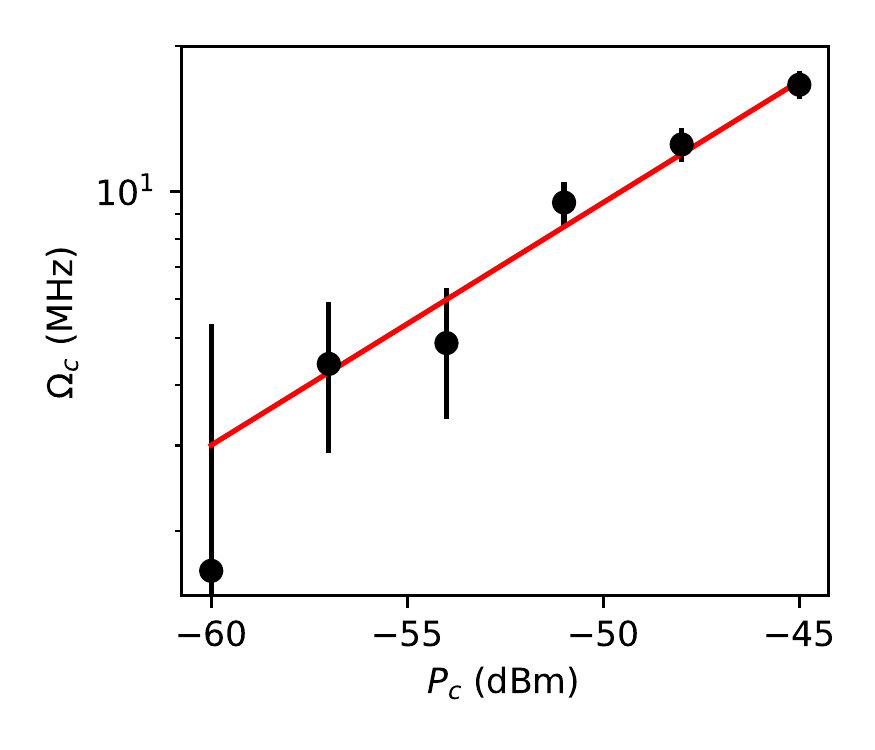}
\caption{Effective control tone strength $\Omega_c$ obtained for each data point. The solid red line is the $\Omega_c=\sqrt{kP_c}$ drive strength given by the fit result.}
\label{Omega_loglog}
\end{center}
\end{figure}

The linewidth of the transparency window when sweeping the control field is given by eq.~\ref{eq: linewidth} as $\gamma_\mathrm{EIT} = \gamma_{20}+\Omega_c^2/(4\gamma_{10})$. The square of the drive strength is proportional to the input power, $\Omega_c^2=k P_c$, where $k$ is related to the attenuation of the microwave lines as well as the coupling capacitance of the qubit to the electrical gate. For the linewidth this implies $\gamma_\mathrm{EIT} = \gamma_{20}+k P_c/(4\gamma_{10})$. A sweep of the control power gives a linearly increasing $\gamma_\mathrm{EIT}$ with $P_c$, with slope $k/(4\gamma_{10})$ and y-intercept $\gamma_{20}$. We determine $\gamma_{10}$ from a measurement without the control field turned on, and then extract the parameters $\gamma_{20}$ and $k$ from a line fit, which in turn yields the drive strength $\Omega_c$. The solid red line in Fig. \ref{linewidths} is given by $\sqrt{kP_c}$. We also calculate $\Omega_c$ for each measured value of $\gamma_\mathrm{EIT}$ along with error bars. At low control power the EIT linewidth is dominated by the intrinsic linewidth $\gamma_{20}$, giving rise to large error margins. In Fig.~\ref{Omega_loglog} we plot the $\Omega_c$ results on a logarithmic scale.

\end{document}